\documentclass[fleqn,twocolumn,amsmath,amssymb,longbibliography]{revtex4-1}

\usepackage{xifthen}
\newcommand{\refAppendix}[6]{#1
  \ifthenelse{\isempty{#2}}%
    {}
    {\protect\cite{#2}}
    #3\protect\ref{#4}#5#6\xspace
  }
  
\usepackage{forloop,ifthen} 

\usepackage{amsfonts}
\usepackage{mathtools}
\usepackage{amsmath}
\usepackage{amssymb}
\usepackage{latexsym}

\usepackage{graphicx}
\usepackage{times}
\usepackage[colorlinks=true,linkcolor=blue,citecolor=red,plainpages=false,pdfpagelabels]{hyperref}
\usepackage{xspace}

\usepackage{color}

\usepackage[normalem]{ulem}

\providecommand{\U}[1]{\protect\rule{.1in}{.1in}}

\usepackage{multirow} 

\begin{document}

\title{Generation of heralded optical `Schr\"odinger cat' states by photon-addition}

\author{Yi-Ru Chen,$^{1}$ Hsien-Yi Hsieh,$^{1}$ Jingyu Ning,$^{1}$ Hsun-Chung Wu,$^{1}$ Hua Li
  Chen,$^{2}$ Zi-Hao Shi,$^{1}$ Popo Yang,$^{1}$ Ole Steuernagel,$^{1}$ Chien-Ming Wu,$^{1}$ and
  Ray-Kuang Lee$^{1,2,3,4}$}
 \email{rklee@ee.nthu.edu.tw}

\affiliation{$^{1}$Institute of Photonics Technologies, National Tsing Hua University, Hsinchu 30013, Taiwan\\
$^{2}$Department of Physics, National Tsing Hua University, Hsinchu 30013, Taiwan\\
$^{3}$Physics Division, National Center for Theoretical Sciences, Taipei 10617, Taiwan\\
$^{4}$Center for Quantum Technology, Hsinchu 30013, Taiwan}

\date{\today}
\begin{abstract}
  Optical ‘Schr\"odinger cat’ states, the non-classical superposition of two quasi-classical
  coherent states, serve as a basis for gedanken experiments testing quantum physics on mesoscopic
  scales and are increasingly recognized as a resource for quantum information processing. Here, we
  report the first experimental realization of optical ‘Schr\"odinger cats’ by adding a photon to a
  squeezed vacuum state, so far only photon-subtraction protocols have been
  realized. Photon-addition gives us the advantage of using heralded signal photons as
  experi\-men\-tal triggers, and we can generate ‘Schr\"odinger cats’ at rates exceeding
  $8.5 \times 10^4$ counts per second; at least one order of magnitude higher than all previously
  reported realizations. Wigner distributions with pronounced nega\-tive parts are demonstrated at
  down to -8.89~dB squee\-zing, even when the initial squeezed vacuum input state has low
  purity. Benchmarking against such a degraded squeezed input state we report a maximum fidelity of
  more than $80\%$ with a maximum cat amplitude of $|\alpha| \approx 1.66$. Our experiment uses
  photon-addition from pairs, one of those photons is used for monitoring, giving us enhanced
  control; moreover the pair production rates are high and should allow for repeated application of
  photon-addition via repeat-stages.
\end{abstract}
\maketitle

Coherent laser fields described by the quantum state $|\alpha\rangle$ can be put into a mesoscopic
superposition of two such states $|\alpha\rangle - |-\alpha\rangle$~\cite{theory-cs,
  theory-entangle, theory-ss}. Here the light field points in opposite direc\-tions $\pm \alpha$ at
the same time, and this is superposed in such a way that the relative phase between the individual
components leads to destructive interference at the center.  Such super\-position states are called
optical Schr\"odinger ``cat states'', in honor of Erwin Schr\"odinger's famous {\it gedanken
  experiments} to illustrate paradoxa of quantum physics~\cite{Sch-1}.

Such states are important as tools for testing macroscopic limits of quantum physics and also for
many quantum computation and communication protocols~\cite{CV, Gaussian}. But they are hard to
generate and so far only available at low production rate and low amplitudes,~$\alpha$, when using
photon-subtraction schemes~\cite{kitten}.

To go beyond these constrains, here, we report the first experi\-mental realization of
photon-addition for the generation of `negative cat' mesoscopic superposition states, of the form
\begin{equation}
  \label{eq:catState}
\Psi_-  =  N_\alpha (|\alpha\rangle - |-\alpha\rangle ) \; ,
\end{equation}
from highly squeezed input states. $N_\alpha$ is the normalization constant.

Traditionally, the constituent states~$|\alpha\rangle$ are described as weakly excited Glauber
coherent states but in our case the input states can be so highly squeezed, that we provide an
additional modified description to determine the quantum fidelity of the performance of our setup.
We do this by benchmarking against experimentally determined input states which are subjected to
photon addition,~$\hat \rho_{\rm sq}^{\rm add}$, as well as Glauber state superpositions
$\Psi_-$~(\ref{eq:catState}).

Our approach, using photon-addition, has three immediate advantages over the traditional
photon-subtraction approach: the generation rate increases by an order of magnitude, the states are
non-classical even for high initial squee\-zing and low initial state purity, and the
amplitude~$|\alpha|$ of the final state, very desirably, increases when using photon-addition,
instead of decreasing it, as is unavoidable when using subtraction.

The earliest theoretical proposal for photon-addition to coherent states~\cite{add-CS} was soon
followed by those for addition to a squeezed vacuum state~\cite{PACS, PASVS} and an experimental
demonstration of photon-addition and the quantum-to-classical transition for coherent states of
light~\cite{add-trans}. Also, quantum commutation rules for thermal states were probed
experimentally~\cite{add-sub-therm, add-therm}. Here, we perform the first addition of a single
photon to a squeezed vacuum state, both constituent states have no classical counterpart.

Using an optical parametric oscillator (OPO) in the continuous wave regime, photon-subtraction from
a squeezed vacuum state has been implemented for the generation of `Schr\"odinger
cats'~\cite{odd, OE-07}. Production of cat states based on controlled removal of a single
photon has to use weakly reflecting mirrors, to avoid damage to the desired output-states. This
badly limits the success rate of such subtraction approaches.

To boost the state's usefulness for fundamental and practical applications, several schemes have
been attempted to enlarge the size of cat states, such as using two-photon Fock
states~\cite{cat-07}, ancilla-assisted photon-subtraction~\cite{lancilla}, and optical
synthesis~\cite{synthesis}.  Even though high purity and entanglement of optical cat
states~\cite{cat-filter, cat-enlarge}, as well as remote preparation of non-local
superpositions~\cite{cat-remote}, have been demonstrated, a high generation rate is crucial for
practical applications of such superpositions in fundamental tests and quantum information
protocols.

Contrasted with photon-subtraction, photon-addition from an entangled photon-pair allows us to use
one of the pair as a heralding photon to monitor success whilst also substantially increasing
production rates.  Therefore, only through photon-addition can we reasonably hope to add more
photons in further stages whilst preserving, or even enhancing further, the non-classical nature of
the final state.

A pure squeezed vacuum state,~$|\xi\rangle$, to which $m$ photons are added, has the form
$|\xi,m\rangle=\frac{1}{\sqrt{N_m(|\xi|)}} (\hat a^\dagger)^m |\xi\rangle$. Here,
$N_m(|\xi|)=m!(1-|\xi|^2)^{-m/2}P_m[(1-|\xi|^2)^{-1/2}]$ is the normali\-zation factor with
Legendre polynomials~$P_m$~\cite{PASVS}.

In the case of one added photon ($m=1$), the single-photon–added pure squeezed vacuum state is
$|\xi,1\rangle=\frac{\hat a^\dagger |\xi\rangle}{\sqrt{N_1(|\xi|)}}$ and only a negative-cat
superposition state~ $\Psi_-$~(\ref{eq:catState}) can result~\cite{addsq-1, addsq-2}. This is
achieved using the signal photon of photon-pairs from spontaneous parametric down conversion
(SPDC), formally
\begin{equation}
  \!\!\!\!\!\!\!\!\!
  |\psi\rangle \approx (1+g\, \hat a_s^\dagger\hat a_i^\dagger)|\xi\rangle_s|0\rangle_i
= |\xi\rangle_s|0\rangle_i+g\, \hat a_s^\dagger|\xi\rangle_s|1\rangle_i \; . \quad
\end{equation}
Here, $g$ is the gain factor of the SPDC process, and $\hat{a}_s^\dagger$ and $\hat{a}_i^\dagger$
are signal and idler photon creation operators, respectively.

\begin{figure}[b]
  \centering
\includegraphics[width=8.4cm]{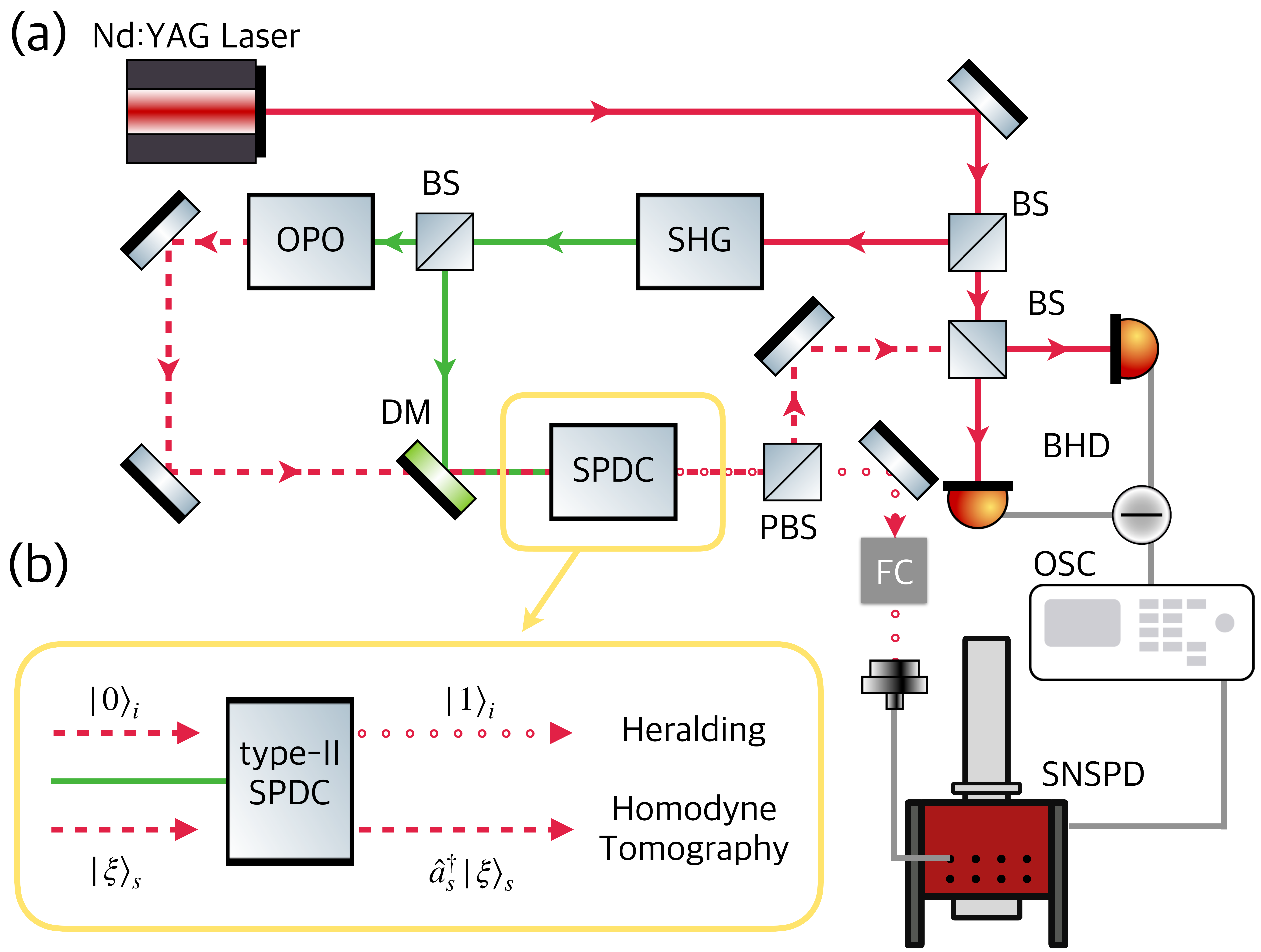}
\caption{ \label{Fig:ExpNew}Schematic diagram of the experimental setup. (a) The squeezed vacuum
  beam $|\xi\rangle_s$ is produced by the OPO (optical parametric oscillator) and injected into the
  type-II SPDC (spontaneous parametric down conversion) to perform the photon-addition scheme
  $\hat{a}_s^\dagger |\xi\rangle_s$.  (b) After passing through the filter cavity (FC), the idler
  single photon state $|1\rangle_i$ is used as the heralding signal, collected by the
  superconducting nanowire single photon detector (SNSPD). The optical `Schr\"odinger cats', i.e.,
  photon-added squeezed state, is characterized by homodyne tomography with the balanced
  homodyne detectors (BHD). The setup also includes BS (beam splitter), DM (dichroic mirrors), PBS
  (polarization beam splitter), SHG (second-harmonic generator), and OSC (oscilloscope).}
\end{figure}

In our experiment, the squeezed vacuum state, $\varrho_{\rm sq}^{\rm OPO}$, at the output of the
OPO, see Fig.~\ref{Fig:ExpNew}, is degraded. We reach squee\-zing levels as low as -8.89~dB, with a
purity of 0.49. Yet, after addition of a single photon we reach a `relative
fidelity',~$F_{\text{sq}}^{\text{add}}$, of 80.1\% with a maximum cat size $|\alpha| \approx$ 1.66
at a generation rate of 8.5 $ \times 10^4$ counts per second, this is at least one order of
magnitude higher than all previously reported rates (which are all based on photon-subtraction).

\begin{figure*}[t!]
  \centering
\includegraphics[width=16.8cm]{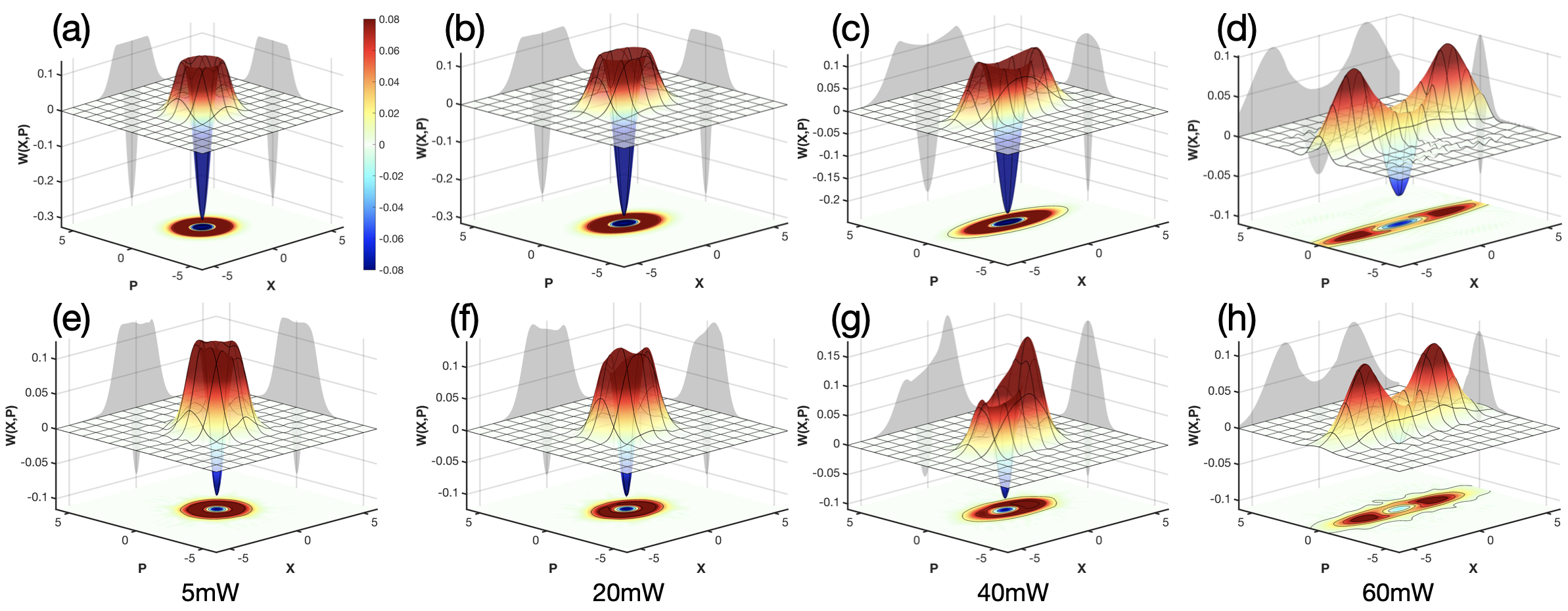}
\caption{ \label{fig:WignerCats}Wigner distributions of optical Schr\"odinger cats: (a)-(d) for the
  theoretical photon-added squeezed states, $W[\hat \rho_{\text{sq}}^{\text{add}}]$ of
  Eq.~(\ref{addsq}); and (e)-(h) for experimental data, $W[\varrho_{\text{sq}}^{\text{add}}]$,
  obtained with OPO pump-powers at $P_{\text{OPO}} = 5$, $20$, $40$, and $60$ mW, respectively. }
\end{figure*}

\vspace{0.2in}
\begin{table*}[t]
  \begin{tabular}{| c  | c | c | c | c |  c | c | c | c | c| }
    \hline
    $P_{\text{OPO}}$ ($P_{\text{rel}}$\%)  & SQ~:~ASQ ($\varrho_{\text{sq}}^{\text{OPO}} $)
    & Purity ($\varrho_{\text{sq}}^{\text{OPO}} $) &SQ~:~ASQ ($\varrho_{\text{sq}}^{\text{empty}} $)
    &  Purity ($\varrho_{\text{sq}}^{\text{add}}$)
    &  $W[\varrho_{\text{sq}}^{\text{add}}](0,0)$  & $F_\text{cat}$ &  $F_\text{sq}^\text{add}$ &  Cat size ($|\alpha|$) & $\Gamma$(counts/s)\\
    \hline

5 \text{mW} (5\%)& -3.76~:~3.89 dB& 0.99& -0.33~:~0.50 dB& 0.54&  -0.095 & 64.9\%& 66.3\% &	0.28& 	4.4 $ \times  10^4$ \\
20 \text{mW} (19\%)& -6.27~:~7.31 dB &  	0.89&  -1.42~:~1.78 dB& 0.55& -0.104& 65.9\%& 68.7\% &	0.62& 	4.5 $ \times  10^4$\\
40 \text{mW} (38\%)& -7.59~:~11.55 dB& 	0.63& -3.42~:~4.87 dB&  0.54& -0.090& 61.8\%& 68.7\% & 	1.02&  	5.1 $ \times  10^4$\\
60 \text{mW} (58\%)&	-8.89~:~15.13 dB&  0.49& -4.17~:~9.85 dB&   0.30& -0.023& 42.4\%&  80.1\% & 	1.66& 	8.5 $ \times  10^4$\\
    \hline
         \end{tabular}
         \caption{\label{table1} Results summary for a fixed SPDC pump-power of 10~mW at different
           OPO pump-power levels~$P_{\text{OPO}}$ (OPO threshold power of $P_{\text{th}}=$ 103.9~mW,
           hence, power relative to threshold $P_{\text{rel}} = P_{\text{OPO}} / P_{\text{th}}$
           in~\%), column 1.  Columns 2 and 3, squee\-zing (SQ) and anti-squee\-zing (ASQ) levels for
           state~$\varrho_{\text{sq}}^{\text{OPO}}$ right after the OPO and its purity. Column 4,
           SQ~:~ASQ values obtained after the empty SPDC cavity (without pump field, i.e., for
           $\varrho_{\text{sq}}^{\text{empty}}$). Column 5, purity of the output-state,
           $\varrho_{\text{sq}}^{\text{add}}$, after performing the photon-addition and, column 6,
           its negative Wigner distribution values
           $W[\varrho_{\text{sq}}^{\text{add}}](0,0)$. Fidelities w.r.t. 
           Glauber cat states~(\ref{eq:catState}) and photon-added impure squeezed
           states~(\ref{addsq}), 
           columns 7 and 8. Cat size $|\alpha|$, and generation rates $\Gamma$, columns 9 and 10. All
           squee\-zing data are averaged over $10$ measurements with a statistical error of~0.12~dB.
         }
\end{table*}

As sketched in Fig.~\ref{Fig:ExpNew}~(a), we initially generate squeezed vacuum states in a bow-tie
OPO cavity enclosing a periodi\-cally poled nonlinear KTiOPO$_4$ (PPKTP) crystal with second order
nonlinear susceptibility,~$\chi^{(2)}$, operated below the lasing threshold, at wavelength
1064~nm~\cite{CMRR}. This OPO-cavity has an optical path length of approximately $285$~mm and thus a
free spectral range of $1.052$ GHz, with a finesse of $19.8$ at $532$ nm and of $33.4$ at $1064$
nm. The overall efficiency,~$1-L$ (where $L$ is the loss), is $82.2 \pm 0.35\%$ and the
phase noise is $34.50 \pm 1.26$ mrad.

The OPO-generated squeezed vacuum states are subsequently injected into a type-II SPDC-crystal which
is inside another bow-tie cavity of the same configuration as the OPO-cavity, i.e., also $285$ mm in
length, with full-width at half maximum (FWHM) of 31.8~MHz. It performs the photon-addition
[formally $\hat{a}_s^\dagger |\xi\rangle$, compare Fig.~\ref{Fig:ExpNew}~(b)].

The OPO-cavity's free spectral range is $1.052$ GHz giving rise to neighboring modes which we
suppress with filter cavities, consisting of one triangle cavity followed by two Fabry-Perot
cavities, with FWHM of 16.3, 131, and 101~MHz, respectively. This filtering plays a critical role in
our experiments since it ensures that, at a time, only a single signal-photon is added into the
squeezed vacuum. The SPDC-pair's idler photon gets detected by the superconducting nanowire
single-photon detector, generating the heralding trigger signal for the homodyne detector, the
output of which is used to tomo\-graphically reconstruct the Schr\"odinger cats, see
Fig.~\ref{Fig:ExpNew}.

The magnitudes of squee\-zing (SQ) versus anti-squee\-zing (ASQ), at the output of the
OPO,~$\varrho_{\rm sq}^{\rm OPO}$, are almost the same at a low pump-power of 5~mW,
specifically~\cite{QST-ML}: SQ~:~ASQ = $-3.77$~:~$3.89$ dB. At the highest pump-power of $60$ mW,
the squee\-zing ratio is SQ~:~ASQ = $-8.89$~:~$15.13$ dB.  Simultaneously, the purity of the squeezed
vacuum drops from 0.99 to 0.49, see columns 2 and 3 of Table~\ref{table1} below.

Note, we use the denotation `$\hat \rho$' for theoretical and `$\varrho$' for experimentally
reconstructed quantum states.

The SQ~:~ASQ levels are additionally measured after the SPDC cavity, without applying SPDC pump
power, yielding the quantum state
$\varrho_{\text{sq}}^{\text{empty}} = {\cal L}[\varrho_{\rm sq}^{\rm OPO}] $, where $\cal L$ is a
Lindbladian superoperator describing the effects of the empty SPDC cavity, see column 4 of
Table~\ref{table1}. We do not model these effects explicitly but determine the losses due to the
presence of the empty SPDC-cavity in the experimental reconstruction of
$\varrho_{\text{sq}}^{\text{empty}}$, which subsequently serves as our benchmark for the
quantification of the `relative fidelity',~$F_{\text{sq}}^{\text{add}}$, of the photon-addition process.

At an SPDC pump-power of $10$ mW, we can ensure that the purity of our signal photon (which is to be
added) shows a second-order correlation function of $g^{(2)}(0) \approx 0.55$, confirming that we
avoid detrimental multiple-photon events. This SPDC pump-power of $10$ mW is therefore from now on
chosen and kept fixed while we generate photon-added squeezed states at different OPO pump-powers,
namely at $5$, $20$, $40$, and $60$~mW, see Table~\ref{table1}.  In
Fig.~\ref{fig:WignerCats}~(a)-(d) we display the Wigner
distributions,~$W[\hat \rho_{\text{sq}}^{\text{add}}]$, for the theoretical
description~$\hat \rho_{\text{sq}}^{\text{add}}$ of photon-added impure squeezed states, formally
generated from the expression
\begin{eqnarray}
\label{addsq}
\hat \rho_{\text{sq}}^{\text{add}} = \hat{a}_s^\dagger\, \varrho_{\text{sq}}^{\text{empty}} \, \hat{a}_s \; .
\end{eqnarray}
The calculation of states~(\ref{addsq}) is based on the fact that we use degraded squeezed vacuum
states, $\varrho_{\text{sq}}^{\text{empty}}$ (note that we have dropped explicit reference to state
normalization). Tomo\-graphic reconstructions of the Wigner distributions
$W[\varrho_{\text{sq}}^{\text{add}}]$ (from BHD-data, see Fig.~\ref{Fig:ExpNew}) are shown
in Fig.~\ref{fig:WignerCats}~(e)-(h).

In Fig.~\ref{fig:WignerCats}~(a) and (e), we display the Wigner distribution when using low OPO
pump-power of $P_{\text{OPO}} =$ 5~mW.  The experimentally reconstructed state
$\varrho_{\text{sq}}^{\text{add}}$ has a puri\-ty of $0.54$ and a negative value of
$W[\varrho_{\text{sq}}^{\text{add}}](0,0) = -0.095$.  Comparison between experimental Wigner
distributions and Glauber-cat reference states~$\Psi_-$ of Eq.~(\ref{eq:catState}) give good
agreement at such low pump-power levels, as the degradation in the injected squeezed vacuum states
is not severe. Maximi\-zing the fidelity with respect to an ideal Glauber cat state~$\Psi_-$ allows
us to estimate the amplitude of our cat state as $|\alpha| = 0.28$, see Table~\ref{table1} column~9.

When we increase the OPO pump-power to 20 and 40~mW, the OPO output
state,~$\varrho_{\text{sq}}^{\text{OPO}}$, starts to show degraded squee\-zing levels, SQ~:~ASQ =
-6.27~:~7.31~dB and -7.59~:~11.55~dB, see Table~\ref{table1} column~2.  Yet, despite this degradation,
the purity in the target cat states remains as high as that from the nearly perfectly pure squeezed
state, i.e., 0.55 and 0.54, respectively, see Table~\ref{table1} column~7.  For both cases, the
negativity remains pronounced and the amplitude of our cat states enlarges to $|\alpha| = 0.62$ and
$1.02$, respectively. See Table~\ref{table1} columns 6 and~9, and Fig.~\ref{fig:WignerCats}~(f),~(g)
for the associated Wigner distributions~$W[\varrho_{\text{sq}}^{\text{add}}]$.

Finally, we reach an OPO pump-power of 60~mW. Now the OPO
output-state,~$\varrho_{\text{sq}}^{\text{OPO}}$, shows further degraded squee\-zing levels and a
low purity of $0.49$, see Table~\ref{table1} column~3.  Therefore, the final optical Schr\"odinger
cat state,~$\varrho_{\text{sq}}^{\text{add}}$, also has a low purity of $0.30$, see
Table~\ref{table1} column~5.  The associated `cat amplitude' is large, $|\alpha| = 1.66$, see
Table~\ref{table1} column~9 and Fig.~\ref{fig:WignerCats}~(h):

Most surprisingly, however, despite the input state~$\varrho_{\text{sq}}^{\text{OPO}}$'s low purity (at
60~mW OPO pump-power) and a noisy environment, using photon-addition lets the optical cat `survive':
it shows a clear negative value of~ $W[\varrho_{\text{sq}}^{\text{add}}](0,0) = -0.023$.

Overall, we observe a maximum negativity of $-0.104$ at an OPO pump-power of $20$ mW when the purity of our
generated optical cat stats also reaches its largest value.

It is known that, when applying photon-subtraction, the input squeezed states must be pure enough,
practically limiting the input state's squee\-zing to no less than $-5$ dB.
Additionally, repeated application of photon-subtraction drives states towards the vacuum
state~\cite{PhysRevA.82.043842}.

By contrast, our photon-addition scheme can support the generation of optical cat states with higher
squee\-zing levels, even as the input squeezed states become increasingly impure.

Photon-addition is so tolerant to experimental imperfections that we still observe pronounced
negative parts in the Wigner distribution all the way down to $-8.89$ dB in squee\-zing, even though
the input squeezed state has low purity, see Table~\ref{table1} column~6.

We compare our experimentally reconstructed state $\varrho_{\text{sq}}^{\text{add}}$ with the ideal
Glauber cat state ($|\alpha\rangle - |-\alpha\rangle$) by calculating its fidelity
$F_\text{cat}$. This fidelity remains higher than $60\%$ even when working at the OPO pump-power of
40~mW, but at 60~mW pump-power it drops to $42.4\%$, see Table~\ref{table1}~column~7. We will now
explain that this drop in fidelity is a poor measure of the good performance of photon-addition; the
latter is better quantified by the relative fidelity~$F_{\text{sq}}^{\text{add}}$ with respect to
state~(\ref{addsq}), see Table~\ref{table1} column~8:

When we introduced the negative cat state~$\Psi_-$~of~Eq.~(\ref{eq:catState}), based on Glauber
states, we emphasized that our squee\-zing levels are so high and therefore, at the current state of
technology for the generation of squeezed states, the impurity so large, that descriptions by
Glauber states as constituents become ill-matched. To address this shortcoming in the
\emph{theoretical analysis} we therefore directly take into account the degradation in the injected
squeezed states by determining it experimentally as
$\varrho_\text{sq}^\text{empty}$~\cite{QST-ML}. We then extract the fidelity due to photon-addition
by using the \emph{experimentally reconstructed} density matrix $\varrho_{\text{sq}}^{\text{add}}$
and comparing it to the ideal photon-addition upon $\varrho_\text{sq}^\text{empty}$, namely by
comparison with the state~ $\hat \rho_{\text{sq}}^{\text{add}}$ of Eq.~(\ref{addsq}). This, gives us
the benchmark for the extraction of the relative fidelity,~$F_{\text{sq}}^{\text{add}}$, of only the
photon-addition operation by itself.

To our delight, this analysis confirms that photon-addition is robust. Adding a photon to a highly
squeezed impure input state has a high relative fidelity~$F_{\text{sq}}^{\text{add}}$-value of
$80.1\%$ at 60~mW, see Table~\ref{table1}~column~8.

We tentatively attribute this good performance to the fact that the highly squeezed impure input
state $\varrho_\text{sq}^\text{empty}$ has broad support in phase space and thus lower peak values
than when it is generated at lower OPO pump-power, see Fig.~\ref{fig:WignerCats}~(d) and~(h). This
should make it easier for the added photon to `punch a negative hole' into its Wigner distribution,
thus increasing the relative fidelity.

These interpretations are in agreement with our finding that the `cat survives' to extremely high
levels of squee\-zing whilst being generated from a rather impure squeezed input state.

We have shown that photon-addition can be applied to a highly squeezed input state with a
squee\-zing level, SQ, of $-8.89$~dB. This corresponds to a cat state with an amplitude of 1.66 and
a fidelity for the photon-addition process of $80.1\%$ whilst maintaining control through the use of
idler photons at a generation rate of 8.5 $\times 10^4$ per second, this is at least one order of
magnitude higher than all previously reported rates.

Our experiments thus overturn the common belief that sequential application of photon-addition
through multiple OPOs would involve significant losses and is therefore not advisable in
experimental practice~\cite{review}.

Instead, our successful generation of heralded optical Schr\"odinger cats by photon-addition opens
up the possibility towards generation of still larger cat states by repeatedly adding
photons~\cite{add-cat}.

We emphasize that imperfections of our experiment are to a considerable extent due to the fact that
the squeezed states we start from have reduced levels of purity when they are highly squeezed. In
other words, the synthesis of highly squeezed states with high purity remains a desirable and
partially unfulfilled goal of the quantum optics community.  By contrast, we establish here that the
addition of a photon can already be controlled to a surprisingly good degree, we quantified this
using a well-matched `relative fidelity' measure.

Controlled photon-addition promises to become a new powerful building block for advanced quantum
state synthesis.  With such a high generation rate, our photon-addition approach also promises to
facilitate applications in quantum information processing using cat codes~\cite{catcode, catqubit},
or preparing error-correcting codes~\cite{GKP, OL, blueprint, catalysis, approx}.

This work is partially supported by the Ministry of Science and Technology of Taiwan
(Nos. 108-2923-M-007-001-MY3 and 110-2123-M-007-002), Office of Naval Research Global, US Army
Research Office (ARO), and the collaborative research program of the Institute for Cosmic Ray
Research (ICRR), the University of Tokyo.

\bibliography{biblio_doi}

\begin{thebibliography}{35}%
\makeatletter
\providecommand \@ifxundefined [1]{%
 \@ifx{#1\undefined}
}%
\providecommand \@ifnum [1]{%
 \ifnum #1\expandafter \@firstoftwo
 \else \expandafter \@secondoftwo
 \fi
}%
\providecommand \@ifx [1]{%
 \ifx #1\expandafter \@firstoftwo
 \else \expandafter \@secondoftwo
 \fi
}%
\providecommand \natexlab [1]{#1}%
\providecommand \enquote  [1]{``#1''}%
\providecommand \bibnamefont  [1]{#1}%
\providecommand \bibfnamefont [1]{#1}%
\providecommand \citenamefont [1]{#1}%
\providecommand \href@noop [0]{\@secondoftwo}%
\providecommand \href [0]{\begingroup \@sanitize@url \@href}%
\providecommand \@href[1]{\@@startlink{#1}\@@href}%
\providecommand \@@href[1]{\endgroup#1\@@endlink}%
\providecommand \@sanitize@url [0]{\catcode `\\12\catcode `\$12\catcode
  `\&12\catcode `\#12\catcode `\^12\catcode `\_12\catcode `\%12\relax}%
\providecommand \@@startlink[1]{}%
\providecommand \@@endlink[0]{}%
\providecommand \url  [0]{\begingroup\@sanitize@url \@url }%
\providecommand \@url [1]{\endgroup\@href {#1}{\urlprefix }}%
\providecommand \urlprefix  [0]{URL }%
\providecommand \Eprint [0]{\href }%
\providecommand \doibase [0]{http://dx.doi.org/}%
\providecommand \selectlanguage [0]{\@gobble}%
\providecommand \bibinfo  [0]{\@secondoftwo}%
\providecommand \bibfield  [0]{\@secondoftwo}%
\providecommand \translation [1]{[#1]}%
\providecommand \BibitemOpen [0]{}%
\providecommand \bibitemStop [0]{}%
\providecommand \bibitemNoStop [0]{.\EOS\space}%
\providecommand \EOS [0]{\spacefactor3000\relax}%
\providecommand \BibitemShut  [1]{\csname bibitem#1\endcsname}%
\let\auto@bib@innerbib\@empty
\bibitem [{\citenamefont {Marek}\ \emph {et~al.}(2008)\citenamefont {Marek},
  \citenamefont {Jeong},\ and\ \citenamefont {Kim}}]{theory-cs}%
  \BibitemOpen
  \bibfield  {author} {\bibinfo {author} {\bibfnamefont {P.}~\bibnamefont
  {Marek}}, \bibinfo {author} {\bibfnamefont {H.}~\bibnamefont {Jeong}}, \ and\
  \bibinfo {author} {\bibfnamefont {M.~S.}\ \bibnamefont {Kim}},\ }\bibfield
  {title} {\enquote {\bibinfo {title} {Generating ``squeezed" superpositions of
  coherent states using photon addition and subtraction},}\ }\href {\doibase
  10.1103/PhysRevA.78.063811} {\bibfield  {journal} {\bibinfo  {journal} {Phys.
  Rev. A}\ }\textbf {\bibinfo {volume} {78}},\ \bibinfo {pages} {063811}
  (\bibinfo {year} {2008})}\BibitemShut {NoStop}%
\bibitem [{\citenamefont {Navarrete-Benlloch}\ \emph
  {et~al.}(2012)\citenamefont {Navarrete-Benlloch}, \citenamefont
  {Garcia-Patron}, \citenamefont {Shapiro},\ and\ \citenamefont
  {Cerf}}]{theory-entangle}%
  \BibitemOpen
  \bibfield  {author} {\bibinfo {author} {\bibfnamefont {C.}~\bibnamefont
  {Navarrete-Benlloch}}, \bibinfo {author} {\bibfnamefont {R.}~\bibnamefont
  {Garcia-Patron}}, \bibinfo {author} {\bibfnamefont {J.~H.}\ \bibnamefont
  {Shapiro}}, \ and\ \bibinfo {author} {\bibfnamefont {N.~J.}\ \bibnamefont
  {Cerf}},\ }\bibfield  {title} {\enquote {\bibinfo {title} {Enhancing quantum
  entanglement by photon addition and subtraction},}\ }\href {\doibase
  10.1103/PhysRevA.86.012328} {\bibfield  {journal} {\bibinfo  {journal} {Phys.
  Rev. A}\ }\textbf {\bibinfo {volume} {86}},\ \bibinfo {pages} {012328}
  (\bibinfo {year} {2012})}\BibitemShut {NoStop}%
\bibitem [{\citenamefont {Chabaud}\ \emph {et~al.}(2017)\citenamefont
  {Chabaud}, \citenamefont {Douce}, \citenamefont {Markham}, \citenamefont {van
  Loock}, \citenamefont {Kashefi},\ and\ \citenamefont {Ferrini}}]{theory-ss}%
  \BibitemOpen
  \bibfield  {author} {\bibinfo {author} {\bibfnamefont {U.}~\bibnamefont
  {Chabaud}}, \bibinfo {author} {\bibfnamefont {T.}~\bibnamefont {Douce}},
  \bibinfo {author} {\bibfnamefont {D.}~\bibnamefont {Markham}}, \bibinfo
  {author} {\bibfnamefont {P.}~\bibnamefont {van Loock}}, \bibinfo {author}
  {\bibfnamefont {E.}~\bibnamefont {Kashefi}}, \ and\ \bibinfo {author}
  {\bibfnamefont {G.}~\bibnamefont {Ferrini}},\ }\bibfield  {title} {\enquote
  {\bibinfo {title} {Continuous-variable sampling from photon-added or
  photon-subtracted squeezed states},}\ }\href {\doibase
  10.1103/PhysRevA.96.062307} {\bibfield  {journal} {\bibinfo  {journal} {Phys.
  Rev. A}\ }\textbf {\bibinfo {volume} {96}},\ \bibinfo {pages} {062307}
  (\bibinfo {year} {2017})}\BibitemShut {NoStop}%
\bibitem [{\citenamefont {Schr\"odinger}(1935)}]{Sch-1}%
  \BibitemOpen
  \bibfield  {author} {\bibinfo {author} {\bibfnamefont {E.}~\bibnamefont
  {Schr\"odinger}},\ }\bibfield  {title} {\enquote {\bibinfo {title} {Die
  gegenw\"artige situation in der quantenmechanik (the present situation in
  quantum mechanics)},}\ }\href@noop {} {\bibfield  {journal} {\bibinfo
  {journal} {Naturwissenschaften}\ }\textbf {\bibinfo {volume} {23}},\ \bibinfo
  {pages} {807--812} (\bibinfo {year} {1935})}\BibitemShut {NoStop}%
\bibitem [{\citenamefont {Braunstein}\ and\ \citenamefont {van
  Loock}(2005)}]{CV}%
  \BibitemOpen
  \bibfield  {author} {\bibinfo {author} {\bibfnamefont {S.~L.}\ \bibnamefont
  {Braunstein}}\ and\ \bibinfo {author} {\bibfnamefont {P.}~\bibnamefont {van
  Loock}},\ }\bibfield  {title} {\enquote {\bibinfo {title} {Quantum
  information with continuous variables},}\ }\href {\doibase
  10.1103/RevModPhys.77.513} {\bibfield  {journal} {\bibinfo  {journal} {Rev.
  Mod. Phys.}\ }\textbf {\bibinfo {volume} {77}},\ \bibinfo {pages} {513}
  (\bibinfo {year} {2005})}\BibitemShut {NoStop}%
\bibitem [{\citenamefont {Weedbrook}\ \emph {et~al.}(2012)\citenamefont
  {Weedbrook}, \citenamefont {Pirandola}, \citenamefont
  {Garc{\'\i}a-Patr{\'o}n}, \citenamefont {Cerf}, \citenamefont {Ralph},
  \citenamefont {Shapiro},\ and\ \citenamefont {Lloyd}}]{Gaussian}%
  \BibitemOpen
  \bibfield  {author} {\bibinfo {author} {\bibfnamefont {C.}~\bibnamefont
  {Weedbrook}}, \bibinfo {author} {\bibfnamefont {S.}~\bibnamefont
  {Pirandola}}, \bibinfo {author} {\bibfnamefont {R.}~\bibnamefont
  {Garc{\'\i}a-Patr{\'o}n}}, \bibinfo {author} {\bibfnamefont {N.~J.}\
  \bibnamefont {Cerf}}, \bibinfo {author} {\bibfnamefont {T.~C.}\ \bibnamefont
  {Ralph}}, \bibinfo {author} {\bibfnamefont {J.~H.}\ \bibnamefont {Shapiro}},
  \ and\ \bibinfo {author} {\bibfnamefont {S.}~\bibnamefont {Lloyd}},\
  }\bibfield  {title} {\enquote {\bibinfo {title} {Gaussian quantum
  information},}\ }\href {\doibase 10.1103/RevModPhys.84.621} {\bibfield
  {journal} {\bibinfo  {journal} {Rev. Mod. Phys.}\ }\textbf {\bibinfo {volume}
  {84}},\ \bibinfo {pages} {621} (\bibinfo {year} {2012})}\BibitemShut
  {NoStop}%
\bibitem [{\citenamefont {Ourjoumtsev}\ \emph {et~al.}(2006)\citenamefont
  {Ourjoumtsev}, \citenamefont {Tualle-Brouri}, \citenamefont {Laurat},\ and\
  \citenamefont {Grangier}}]{kitten}%
  \BibitemOpen
  \bibfield  {author} {\bibinfo {author} {\bibfnamefont {A.}~\bibnamefont
  {Ourjoumtsev}}, \bibinfo {author} {\bibfnamefont {R.}~\bibnamefont
  {Tualle-Brouri}}, \bibinfo {author} {\bibfnamefont {J.}~\bibnamefont
  {Laurat}}, \ and\ \bibinfo {author} {\bibfnamefont {P.}~\bibnamefont
  {Grangier}},\ }\bibfield  {title} {\enquote {\bibinfo {title} {Generating
  optical schr\"odinger kittens for quantum information processing},}\
  }\href@noop {} {\bibfield  {journal} {\bibinfo  {journal} {Science}\ }\textbf
  {\bibinfo {volume} {312}},\ \bibinfo {pages} {83} (\bibinfo {year}
  {2006})}\BibitemShut {NoStop}%
\bibitem [{\citenamefont {Agarwal}\ and\ \citenamefont {Tara}(1991)}]{add-CS}%
  \BibitemOpen
  \bibfield  {author} {\bibinfo {author} {\bibfnamefont {G.~S.}\ \bibnamefont
  {Agarwal}}\ and\ \bibinfo {author} {\bibfnamefont {K.}~\bibnamefont {Tara}},\
  }\bibfield  {title} {\enquote {\bibinfo {title} {Nonclassical properties of
  states generated by the excitations on a coherent state},}\ }\href {\doibase
  10.1007/978-1-4615-2936-1_25} {\bibfield  {journal} {\bibinfo  {journal}
  {Phys. Rev. A}\ }\textbf {\bibinfo {volume} {43}},\ \bibinfo {pages} {492}
  (\bibinfo {year} {1991})}\BibitemShut {NoStop}%
\bibitem [{\citenamefont {Quesne}(2001)}]{PACS}%
  \BibitemOpen
  \bibfield  {author} {\bibinfo {author} {\bibfnamefont {C.}~\bibnamefont
  {Quesne}},\ }\bibfield  {title} {\enquote {\bibinfo {title} {Completeness of
  photon-added squeezed vacuum and one-photon states and of photon-added
  coherent states on a circle},}\ }\href {\doibase
  10.1016/S0375-9601(01)00554-0} {\bibfield  {journal} {\bibinfo  {journal}
  {Phys. Lett. A}\ }\textbf {\bibinfo {volume} {288}},\ \bibinfo {pages} {241}
  (\bibinfo {year} {2001})}\BibitemShut {NoStop}%
\bibitem [{\citenamefont {Zhang}\ and\ \citenamefont {Fan}(1992)}]{PASVS}%
  \BibitemOpen
  \bibfield  {author} {\bibinfo {author} {\bibfnamefont {Z.}~\bibnamefont
  {Zhang}}\ and\ \bibinfo {author} {\bibfnamefont {H.}~\bibnamefont {Fan}},\
  }\bibfield  {title} {\enquote {\bibinfo {title} {Properties of states
  generated by excitations on a squeezed vacuum state},}\ }\href {\doibase
  10.1016/0375-9601(92)91046-T} {\bibfield  {journal} {\bibinfo  {journal}
  {Phys. Lett. A}\ }\textbf {\bibinfo {volume} {165}},\ \bibinfo {pages} {14}
  (\bibinfo {year} {1992})}\BibitemShut {NoStop}%
\bibitem [{\citenamefont {Zavatta}\ \emph {et~al.}(2004)\citenamefont
  {Zavatta}, \citenamefont {Viciani},\ and\ \citenamefont
  {Bellini}}]{add-trans}%
  \BibitemOpen
  \bibfield  {author} {\bibinfo {author} {\bibfnamefont {A.}~\bibnamefont
  {Zavatta}}, \bibinfo {author} {\bibfnamefont {S.}~\bibnamefont {Viciani}}, \
  and\ \bibinfo {author} {\bibfnamefont {M.}~\bibnamefont {Bellini}},\
  }\bibfield  {title} {\enquote {\bibinfo {title} {Quantum-to-classical
  transition with single-photon--added coherent states of light},}\ }\href
  {\doibase 10.1126/science.1103190} {\bibfield  {journal} {\bibinfo  {journal}
  {Science}\ }\textbf {\bibinfo {volume} {306}},\ \bibinfo {pages} {660}
  (\bibinfo {year} {2004})}\BibitemShut {NoStop}%
\bibitem [{\citenamefont {Parigi}\ \emph {et~al.}(2007)\citenamefont {Parigi},
  \citenamefont {Zavatta}, \citenamefont {Kim},\ and\ \citenamefont
  {Bellini}}]{add-sub-therm}%
  \BibitemOpen
  \bibfield  {author} {\bibinfo {author} {\bibfnamefont {V.}~\bibnamefont
  {Parigi}}, \bibinfo {author} {\bibfnamefont {A.}~\bibnamefont {Zavatta}},
  \bibinfo {author} {\bibfnamefont {M.}~\bibnamefont {Kim}}, \ and\ \bibinfo
  {author} {\bibfnamefont {M.}~\bibnamefont {Bellini}},\ }\bibfield  {title}
  {\enquote {\bibinfo {title} {Probing quantum commutation rules by addition
  and subtraction of single photons to/from a light field},}\ }\href {\doibase
  10.1126/science.1146204} {\bibfield  {journal} {\bibinfo  {journal}
  {Science}\ }\textbf {\bibinfo {volume} {317}},\ \bibinfo {pages} {1890}
  (\bibinfo {year} {2007})}\BibitemShut {NoStop}%
\bibitem [{\citenamefont {Zavatta}\ \emph {et~al.}(2007)\citenamefont
  {Zavatta}, \citenamefont {Parigi},\ and\ \citenamefont
  {Bellini}}]{add-therm}%
  \BibitemOpen
  \bibfield  {author} {\bibinfo {author} {\bibfnamefont {A.}~\bibnamefont
  {Zavatta}}, \bibinfo {author} {\bibfnamefont {V.}~\bibnamefont {Parigi}}, \
  and\ \bibinfo {author} {\bibfnamefont {M.}~\bibnamefont {Bellini}},\
  }\bibfield  {title} {\enquote {\bibinfo {title} {Experimental nonclassicality
  of single-photon-added thermal light states},}\ }\href {\doibase
  10.1103/PhysRevA.75.052106} {\bibfield  {journal} {\bibinfo  {journal} {Phys.
  Rev. A}\ }\textbf {\bibinfo {volume} {75}},\ \bibinfo {pages} {052106}
  (\bibinfo {year} {2007})}\BibitemShut {NoStop}%
\bibitem [{\citenamefont {Neergaard-Nielsen}\ \emph {et~al.}(2006)\citenamefont
  {Neergaard-Nielsen}, \citenamefont {Nielsen}, \citenamefont {Hettich},
  \citenamefont {M{\o}lmer},\ and\ \citenamefont {Polzik}}]{odd}%
  \BibitemOpen
  \bibfield  {author} {\bibinfo {author} {\bibfnamefont {J.~S.}\ \bibnamefont
  {Neergaard-Nielsen}}, \bibinfo {author} {\bibfnamefont {B.~Melholt}\
  \bibnamefont {Nielsen}}, \bibinfo {author} {\bibfnamefont {C.}~\bibnamefont
  {Hettich}}, \bibinfo {author} {\bibfnamefont {K.}~\bibnamefont {M{\o}lmer}},
  \ and\ \bibinfo {author} {\bibfnamefont {E.~S.}\ \bibnamefont {Polzik}},\
  }\bibfield  {title} {\enquote {\bibinfo {title} {Generation of a
  superposition of odd photon number states for quantum information
  networks},}\ }\href {\doibase 10.1103/PhysRevLett.97.083604} {\bibfield
  {journal} {\bibinfo  {journal} {Phys. Rev. Lett.}\ }\textbf {\bibinfo
  {volume} {97}},\ \bibinfo {pages} {083604} (\bibinfo {year}
  {2006})}\BibitemShut {NoStop}%
\bibitem [{\citenamefont {Wakui}\ \emph {et~al.}(2007)\citenamefont {Wakui},
  \citenamefont {Takahashi}, \citenamefont {Furusawa},\ and\ \citenamefont
  {Sasaki}}]{OE-07}%
  \BibitemOpen
  \bibfield  {author} {\bibinfo {author} {\bibfnamefont {K.}~\bibnamefont
  {Wakui}}, \bibinfo {author} {\bibfnamefont {H.}~\bibnamefont {Takahashi}},
  \bibinfo {author} {\bibfnamefont {A.}~\bibnamefont {Furusawa}}, \ and\
  \bibinfo {author} {\bibfnamefont {M.}~\bibnamefont {Sasaki}},\ }\bibfield
  {title} {\enquote {\bibinfo {title} {Photon subtracted squeezed states
  generated with periodically poled \text{KTiOPO}$_4$},}\ }\href@noop {}
  {\bibfield  {journal} {\bibinfo  {journal} {Opt. Express}\ }\textbf {\bibinfo
  {volume} {15}},\ \bibinfo {pages} {3568} (\bibinfo {year}
  {2007})}\BibitemShut {NoStop}%
\bibitem [{\citenamefont {Ourjoumtsev}\ \emph {et~al.}(2007)\citenamefont
  {Ourjoumtsev}, \citenamefont {Jeong}, \citenamefont {Tualle-Brouri},\ and\
  \citenamefont {Grangier}}]{cat-07}%
  \BibitemOpen
  \bibfield  {author} {\bibinfo {author} {\bibfnamefont {A.}~\bibnamefont
  {Ourjoumtsev}}, \bibinfo {author} {\bibfnamefont {H.}~\bibnamefont {Jeong}},
  \bibinfo {author} {\bibfnamefont {R.}~\bibnamefont {Tualle-Brouri}}, \ and\
  \bibinfo {author} {\bibfnamefont {P.}~\bibnamefont {Grangier}},\ }\bibfield
  {title} {\enquote {\bibinfo {title} {Generation of optical `schr\"odinger
  cats' from photon number states},}\ }\href@noop {} {\bibfield  {journal}
  {\bibinfo  {journal} {Nature}\ }\textbf {\bibinfo {volume} {448}},\ \bibinfo
  {pages} {784} (\bibinfo {year} {2007})}\BibitemShut {NoStop}%
\bibitem [{\citenamefont {Takahashi}\ \emph {et~al.}(2008)\citenamefont
  {Takahashi}, \citenamefont {Wakui}, \citenamefont {Suzuki}, \citenamefont
  {Takeoka}, \citenamefont {Hayasaka}, \citenamefont {Furusawa},\ and\
  \citenamefont {Sasaki}}]{lancilla}%
  \BibitemOpen
  \bibfield  {author} {\bibinfo {author} {\bibfnamefont {H.}~\bibnamefont
  {Takahashi}}, \bibinfo {author} {\bibfnamefont {K.}~\bibnamefont {Wakui}},
  \bibinfo {author} {\bibfnamefont {S.}~\bibnamefont {Suzuki}}, \bibinfo
  {author} {\bibfnamefont {M.}~\bibnamefont {Takeoka}}, \bibinfo {author}
  {\bibfnamefont {K.}~\bibnamefont {Hayasaka}}, \bibinfo {author}
  {\bibfnamefont {A.}~\bibnamefont {Furusawa}}, \ and\ \bibinfo {author}
  {\bibfnamefont {M.}~\bibnamefont {Sasaki}},\ }\bibfield  {title} {\enquote
  {\bibinfo {title} {Generation of large-amplitude coherent-state superposition
  via ancilla-assisted photon subtraction},}\ }\href {\doibase
  10.1103/PhysRevLett.101.233605} {\bibfield  {journal} {\bibinfo  {journal}
  {Phys. Rev. Lett.}\ }\textbf {\bibinfo {volume} {101}},\ \bibinfo {pages}
  {233605} (\bibinfo {year} {2008})}\BibitemShut {NoStop}%
\bibitem [{\citenamefont {Huang}\ \emph {et~al.}(2015)\citenamefont {Huang},
  \citenamefont {Jeannic}, \citenamefont {Ruaudel}, \citenamefont {Verma},
  \citenamefont {Shaw}, \citenamefont {Marsili}, \citenamefont {Nam},
  \citenamefont {Wu}, \citenamefont {Zeng}, \citenamefont {Jeong},
  \citenamefont {Filip}, \citenamefont {Morin},\ and\ \citenamefont
  {Laurat}}]{synthesis}%
  \BibitemOpen
  \bibfield  {author} {\bibinfo {author} {\bibfnamefont {K.}~\bibnamefont
  {Huang}}, \bibinfo {author} {\bibfnamefont {H.~Le}\ \bibnamefont {Jeannic}},
  \bibinfo {author} {\bibfnamefont {J.}~\bibnamefont {Ruaudel}}, \bibinfo
  {author} {\bibfnamefont {V.~B.}\ \bibnamefont {Verma}}, \bibinfo {author}
  {\bibfnamefont {M.~D.}\ \bibnamefont {Shaw}}, \bibinfo {author}
  {\bibfnamefont {F.}~\bibnamefont {Marsili}}, \bibinfo {author} {\bibfnamefont
  {S.~W.}\ \bibnamefont {Nam}}, \bibinfo {author} {\bibfnamefont
  {E.}~\bibnamefont {Wu}}, \bibinfo {author} {\bibfnamefont {H.}~\bibnamefont
  {Zeng}}, \bibinfo {author} {\bibfnamefont {Y.-C.}\ \bibnamefont {Jeong}},
  \bibinfo {author} {\bibfnamefont {R.}~\bibnamefont {Filip}}, \bibinfo
  {author} {\bibfnamefont {O.}~\bibnamefont {Morin}}, \ and\ \bibinfo {author}
  {\bibfnamefont {J.}~\bibnamefont {Laurat}},\ }\bibfield  {title} {\enquote
  {\bibinfo {title} {Optical synthesis of large-amplitude squeezed
  coherent-state superpositions with minimal resources},}\ }\href {\doibase
  10.1103/PhysRevLett.115.023602} {\bibfield  {journal} {\bibinfo  {journal}
  {Phys. Rev. Lett.}\ }\textbf {\bibinfo {volume} {115}},\ \bibinfo {pages}
  {023602} (\bibinfo {year} {2015})}\BibitemShut {NoStop}%
\bibitem [{\citenamefont {Asavanant}\ \emph {et~al.}(2017)\citenamefont
  {Asavanant}, \citenamefont {Nakashima}, \citenamefont {Shiozawa},
  \citenamefont {Yoshikawa},\ and\ \citenamefont {Furusawa}}]{cat-filter}%
  \BibitemOpen
  \bibfield  {author} {\bibinfo {author} {\bibfnamefont {W.}~\bibnamefont
  {Asavanant}}, \bibinfo {author} {\bibfnamefont {K.}~\bibnamefont
  {Nakashima}}, \bibinfo {author} {\bibfnamefont {Y.}~\bibnamefont {Shiozawa}},
  \bibinfo {author} {\bibfnamefont {J.}~\bibnamefont {Yoshikawa}}, \ and\
  \bibinfo {author} {\bibfnamefont {A.}~\bibnamefont {Furusawa}},\ }\bibfield
  {title} {\enquote {\bibinfo {title} {Generation of highly pure
  schr\"odinger's cat states and real-time quadrature measurements via optical
  filtering},}\ }\href@noop {} {\bibfield  {journal} {\bibinfo  {journal} {Opt.
  Express}\ }\textbf {\bibinfo {volume} {25}},\ \bibinfo {pages} {32227}
  (\bibinfo {year} {2017})}\BibitemShut {NoStop}%
\bibitem [{\citenamefont {Sychev}\ \emph {et~al.}(2017)\citenamefont {Sychev},
  \citenamefont {Ulanov}, \citenamefont {Pushkina}, \citenamefont {Richards},
  \citenamefont {Fedorov},\ and\ \citenamefont {Lvovsky}}]{cat-enlarge}%
  \BibitemOpen
  \bibfield  {author} {\bibinfo {author} {\bibfnamefont {D.~V.}\ \bibnamefont
  {Sychev}}, \bibinfo {author} {\bibfnamefont {A.~E.}\ \bibnamefont {Ulanov}},
  \bibinfo {author} {\bibfnamefont {A.~A.}\ \bibnamefont {Pushkina}}, \bibinfo
  {author} {\bibfnamefont {M.~W.}\ \bibnamefont {Richards}}, \bibinfo {author}
  {\bibfnamefont {I.~A.}\ \bibnamefont {Fedorov}}, \ and\ \bibinfo {author}
  {\bibfnamefont {A.~I.}\ \bibnamefont {Lvovsky}},\ }\bibfield  {title}
  {\enquote {\bibinfo {title} {Enlargement of optical schr{\"o}dinger's cat
  states},}\ }\href@noop {} {\bibfield  {journal} {\bibinfo  {journal} {Nature
  Photon.}\ }\textbf {\bibinfo {volume} {11}},\ \bibinfo {pages} {379}
  (\bibinfo {year} {2017})}\BibitemShut {NoStop}%
\bibitem [{\citenamefont {Ourjoumtsev}\ \emph {et~al.}(2009)\citenamefont
  {Ourjoumtsev}, \citenamefont {Ferreyrol}, \citenamefont {Tualle-Brouri},\
  and\ \citenamefont {Grangier}}]{cat-remote}%
  \BibitemOpen
  \bibfield  {author} {\bibinfo {author} {\bibfnamefont {A.}~\bibnamefont
  {Ourjoumtsev}}, \bibinfo {author} {\bibfnamefont {F.}~\bibnamefont
  {Ferreyrol}}, \bibinfo {author} {\bibfnamefont {R.}~\bibnamefont
  {Tualle-Brouri}}, \ and\ \bibinfo {author} {\bibfnamefont {P.}~\bibnamefont
  {Grangier}},\ }\bibfield  {title} {\enquote {\bibinfo {title} {Preparation of
  non-local superpositions of quasi-classical light states},}\ }\href {\doibase
  10.1038/nphys1199} {\bibfield  {journal} {\bibinfo  {journal} {Nature Phys.}\
  }\textbf {\bibinfo {volume} {5}},\ \bibinfo {pages} {189} (\bibinfo {year}
  {2009})}\BibitemShut {NoStop}%
\bibitem [{\citenamefont {Dakna}\ \emph
  {et~al.}(1998{\natexlab{a}})\citenamefont {Dakna}, \citenamefont
  {Kn{\"o}ll},\ and\ \citenamefont {Welsch}}]{addsq-1}%
  \BibitemOpen
  \bibfield  {author} {\bibinfo {author} {\bibfnamefont {M.}~\bibnamefont
  {Dakna}}, \bibinfo {author} {\bibfnamefont {L.}~\bibnamefont {Kn{\"o}ll}}, \
  and\ \bibinfo {author} {\bibfnamefont {D.-G.}\ \bibnamefont {Welsch}},\
  }\bibfield  {title} {\enquote {\bibinfo {title} {Quantum state engineering
  using conditional measurement on a beam splitter},}\ }\href {\doibase
  10.1007/s100530050177} {\bibfield  {journal} {\bibinfo  {journal} {Eur. Phys.
  J. D}\ }\textbf {\bibinfo {volume} {3}},\ \bibinfo {pages} {295} (\bibinfo
  {year} {1998}{\natexlab{a}})}\BibitemShut {NoStop}%
\bibitem [{\citenamefont {Dakna}\ \emph
  {et~al.}(1998{\natexlab{b}})\citenamefont {Dakna}, \citenamefont
  {Kn{\"o}ll},\ and\ \citenamefont {Welsch}}]{addsq-2}%
  \BibitemOpen
  \bibfield  {author} {\bibinfo {author} {\bibfnamefont {M.}~\bibnamefont
  {Dakna}}, \bibinfo {author} {\bibfnamefont {L.}~\bibnamefont {Kn{\"o}ll}}, \
  and\ \bibinfo {author} {\bibfnamefont {D.-G.}\ \bibnamefont {Welsch}},\
  }\bibfield  {title} {\enquote {\bibinfo {title} {Photon-added state
  preparation via conditional measurement on a beam splitter},}\ }\href
  {\doibase 10.1016/S0030-4018(97)00463-X} {\bibfield  {journal} {\bibinfo
  {journal} {Opt. Comm.}\ }\textbf {\bibinfo {volume} {145}},\ \bibinfo {pages}
  {309} (\bibinfo {year} {1998}{\natexlab{b}})}\BibitemShut {NoStop}%
\bibitem [{\citenamefont {Wu}\ \emph {et~al.}(2019)\citenamefont {Wu},
  \citenamefont {Wu}, \citenamefont {Chen}, \citenamefont {Wu},\ and\
  \citenamefont {Lee}}]{CMRR}%
  \BibitemOpen
  \bibfield  {author} {\bibinfo {author} {\bibfnamefont {C.-M.}\ \bibnamefont
  {Wu}}, \bibinfo {author} {\bibfnamefont {S.-R.}\ \bibnamefont {Wu}}, \bibinfo
  {author} {\bibfnamefont {Y.-R.}\ \bibnamefont {Chen}}, \bibinfo {author}
  {\bibfnamefont {H.-C.}\ \bibnamefont {Wu}}, \ and\ \bibinfo {author}
  {\bibfnamefont {R.-K.}\ \bibnamefont {Lee}},\ }\bibfield  {title} {\enquote
  {\bibinfo {title} {Detection of 10~\text{dB} vacuum noise squeezing at
  1064~nm by balanced homodyne detectors with a common mode rejection ratio
  more than 80~\text{dB}},}\ }in\ \href {\doibase
  10.1364/CLEO_AT.2019.JTu2A.38} {\emph {\bibinfo {booktitle} {Conference on
  Lasers and Electro-Optics}}}\ (\bibinfo  {publisher} {OSA Technical (Optica
  Publishing Group paper JTu2A.38},\ \bibinfo {year} {2019})\BibitemShut
  {NoStop}%
\bibitem [{\citenamefont {Hsieh}\ \emph {et~al.}(2022)\citenamefont {Hsieh},
  \citenamefont {Chen}, \citenamefont {Wu}, \citenamefont {Chen}, \citenamefont
  {Ning}, \citenamefont {Huang}, \citenamefont {Wu},\ and\ \citenamefont
  {Lee}}]{QST-ML}%
  \BibitemOpen
  \bibfield  {author} {\bibinfo {author} {\bibfnamefont {H.-Y.}\ \bibnamefont
  {Hsieh}}, \bibinfo {author} {\bibfnamefont {Y.-R.}\ \bibnamefont {Chen}},
  \bibinfo {author} {\bibfnamefont {H.-C.}\ \bibnamefont {Wu}}, \bibinfo
  {author} {\bibfnamefont {H.~L.}\ \bibnamefont {Chen}}, \bibinfo {author}
  {\bibfnamefont {J.}~\bibnamefont {Ning}}, \bibinfo {author} {\bibfnamefont
  {Y.-C.}\ \bibnamefont {Huang}}, \bibinfo {author} {\bibfnamefont {C.-M.}\
  \bibnamefont {Wu}}, \ and\ \bibinfo {author} {\bibfnamefont {R.-K.}\
  \bibnamefont {Lee}},\ }\bibfield  {title} {\enquote {\bibinfo {title}
  {Extract the degradation information in squeezed states with machine
  learning},}\ }\href {\doibase 10.1103/PhysRevLett.128.073604} {\bibfield
  {journal} {\bibinfo  {journal} {Phys. Rev. Lett.}\ }\textbf {\bibinfo
  {volume} {128}},\ \bibinfo {pages} {073604} (\bibinfo {year}
  {2022})}\BibitemShut {NoStop}%
\bibitem [{\citenamefont {Hu}\ \emph {et~al.}(2010)\citenamefont {Hu},
  \citenamefont {Xu}, \citenamefont {Wang},\ and\ \citenamefont
  {Xu}}]{PhysRevA.82.043842}%
  \BibitemOpen
  \bibfield  {author} {\bibinfo {author} {\bibfnamefont {Li-yun}\ \bibnamefont
  {Hu}}, \bibinfo {author} {\bibfnamefont {Xue-xiang}\ \bibnamefont {Xu}},
  \bibinfo {author} {\bibfnamefont {Zi-sheng}\ \bibnamefont {Wang}}, \ and\
  \bibinfo {author} {\bibfnamefont {Xue-fen}\ \bibnamefont {Xu}},\ }\bibfield
  {title} {\enquote {\bibinfo {title} {Photon-subtracted squeezed thermal
  state: Nonclassicality and decoherence},}\ }\href {\doibase
  10.1103/PhysRevA.82.043842} {\bibfield  {journal} {\bibinfo  {journal} {Phys.
  Rev. A}\ }\textbf {\bibinfo {volume} {82}},\ \bibinfo {pages} {043842}
  (\bibinfo {year} {2010})}\BibitemShut {NoStop}%
\bibitem [{\citenamefont {Lvovsky}\ \emph {et~al.}(2020)\citenamefont
  {Lvovsky}, \citenamefont {Grangier}, \citenamefont {Ourjoumtsev},
  \citenamefont {Parigi}, \citenamefont {Sasaki},\ and\ \citenamefont
  {Tualle-Brouri}}]{review}%
  \BibitemOpen
  \bibfield  {author} {\bibinfo {author} {\bibfnamefont {A.~I.}\ \bibnamefont
  {Lvovsky}}, \bibinfo {author} {\bibfnamefont {P.}~\bibnamefont {Grangier}},
  \bibinfo {author} {\bibfnamefont {A.}~\bibnamefont {Ourjoumtsev}}, \bibinfo
  {author} {\bibfnamefont {V.}~\bibnamefont {Parigi}}, \bibinfo {author}
  {\bibfnamefont {M.}~\bibnamefont {Sasaki}}, \ and\ \bibinfo {author}
  {\bibfnamefont {R.}~\bibnamefont {Tualle-Brouri}},\ }\href@noop {} {\enquote
  {\bibinfo {title} {Production and applications of non-gaussian quantum states
  of light},}\ } (\bibinfo {year} {2020}),\ \Eprint
  {http://arxiv.org/abs/2006.16985} {arXiv:2006.16985} \BibitemShut {NoStop}%
\bibitem [{\citenamefont {{Arman}}\ \emph {et~al.}(2021)\citenamefont
  {{Arman}}, \citenamefont {{Tyagi}},\ and\ \citenamefont
  {{Panigrahi}}}]{add-cat}%
  \BibitemOpen
  \bibfield  {author} {\bibinfo {author} {\bibnamefont {{Arman}}}, \bibinfo
  {author} {\bibfnamefont {Gargi}\ \bibnamefont {{Tyagi}}}, \ and\ \bibinfo
  {author} {\bibfnamefont {Prasanta~K.}\ \bibnamefont {{Panigrahi}}},\
  }\bibfield  {title} {\enquote {\bibinfo {title} {{Photon added cat state:
  phase space structure and statistics}},}\ }\href {\doibase 10.1364/OL.415713}
  {\bibfield  {journal} {\bibinfo  {journal} {Opt. Lett.}\ }\textbf {\bibinfo
  {volume} {46}},\ \bibinfo {pages} {1177} (\bibinfo {year} {2021})},\ \Eprint
  {http://arxiv.org/abs/2011.00990} {arXiv:2011.00990 [quant-ph]} \BibitemShut
  {NoStop}%
\bibitem [{\citenamefont {Chamberland}\ \emph {et~al.}(2022)\citenamefont
  {Chamberland}, \citenamefont {Noh}, \citenamefont {Arrangoiz-Arriola},
  \citenamefont {Campbell}, \citenamefont {Hann}, \citenamefont {Iverson},
  \citenamefont {Putterman}, \citenamefont {Bohdanowicz}, \citenamefont
  {Flammia}, \citenamefont {Keller}, \citenamefont {Refael}, \citenamefont
  {Preskill}, \citenamefont {Jiang}, \citenamefont {Safavi-Naeini},
  \citenamefont {Painter},\ and\ \citenamefont {Brand{\=a}o}}]{catcode}%
  \BibitemOpen
  \bibfield  {author} {\bibinfo {author} {\bibfnamefont {C.}~\bibnamefont
  {Chamberland}}, \bibinfo {author} {\bibfnamefont {K.}~\bibnamefont {Noh}},
  \bibinfo {author} {\bibfnamefont {P.}~\bibnamefont {Arrangoiz-Arriola}},
  \bibinfo {author} {\bibfnamefont {E.~T.}\ \bibnamefont {Campbell}}, \bibinfo
  {author} {\bibfnamefont {C.~T.}\ \bibnamefont {Hann}}, \bibinfo {author}
  {\bibfnamefont {J.}~\bibnamefont {Iverson}}, \bibinfo {author} {\bibfnamefont
  {H.}~\bibnamefont {Putterman}}, \bibinfo {author} {\bibfnamefont {T.~C.}\
  \bibnamefont {Bohdanowicz}}, \bibinfo {author} {\bibfnamefont {S.~T.}\
  \bibnamefont {Flammia}}, \bibinfo {author} {\bibfnamefont {A.}~\bibnamefont
  {Keller}}, \bibinfo {author} {\bibfnamefont {G.}~\bibnamefont {Refael}},
  \bibinfo {author} {\bibfnamefont {J.}~\bibnamefont {Preskill}}, \bibinfo
  {author} {\bibfnamefont {L.}~\bibnamefont {Jiang}}, \bibinfo {author}
  {\bibfnamefont {A.~H.}\ \bibnamefont {Safavi-Naeini}}, \bibinfo {author}
  {\bibfnamefont {O.}~\bibnamefont {Painter}}, \ and\ \bibinfo {author}
  {\bibfnamefont {F.~G. S.~L.}\ \bibnamefont {Brand{\=a}o}},\ }\bibfield
  {title} {\enquote {\bibinfo {title} {Building a fault-tolerant quantum
  computer using concatenated cat codes},}\ }\href@noop {} {\bibfield
  {journal} {\bibinfo  {journal} {PRX Quantum}\ }\textbf {\bibinfo {volume}
  {3}},\ \bibinfo {pages} {010329} (\bibinfo {year} {2022})}\BibitemShut
  {NoStop}%
\bibitem [{\citenamefont {Gravina}\ \emph {et~al.}(2023)\citenamefont
  {Gravina}, \citenamefont {Minganti},\ and\ \citenamefont
  {Savona}}]{catqubit}%
  \BibitemOpen
  \bibfield  {author} {\bibinfo {author} {\bibfnamefont {L.}~\bibnamefont
  {Gravina}}, \bibinfo {author} {\bibfnamefont {F.}~\bibnamefont {Minganti}}, \
  and\ \bibinfo {author} {\bibfnamefont {V.}~\bibnamefont {Savona}},\
  }\bibfield  {title} {\enquote {\bibinfo {title} {Critical schr\"odinger cat
  qubit},}\ }\href@noop {} {\bibfield  {journal} {\bibinfo  {journal} {PRX
  Quantum}\ }\textbf {\bibinfo {volume} {4}},\ \bibinfo {pages} {020337}
  (\bibinfo {year} {2023})}\BibitemShut {NoStop}%
\bibitem [{\citenamefont {Gottesman}\ \emph {et~al.}(2001)\citenamefont
  {Gottesman}, \citenamefont {Kitaev},\ and\ \citenamefont {Preskill}}]{GKP}%
  \BibitemOpen
  \bibfield  {author} {\bibinfo {author} {\bibfnamefont {D.}~\bibnamefont
  {Gottesman}}, \bibinfo {author} {\bibfnamefont {A.}~\bibnamefont {Kitaev}}, \
  and\ \bibinfo {author} {\bibfnamefont {J.}~\bibnamefont {Preskill}},\
  }\bibfield  {title} {\enquote {\bibinfo {title} {Encoding a qubit in an
  oscillator},}\ }\href {\doibase 10.1103/PhysRevA.64.012310} {\bibfield
  {journal} {\bibinfo  {journal} {Phys. Rev. A}\ }\textbf {\bibinfo {volume}
  {64}},\ \bibinfo {pages} {012310} (\bibinfo {year} {2001})}\BibitemShut
  {NoStop}%
\bibitem [{\citenamefont {Vasconcelos}\ \emph {et~al.}(2010)\citenamefont
  {Vasconcelos}, \citenamefont {Sanz},\ and\ \citenamefont {Glancy}}]{OL}%
  \BibitemOpen
  \bibfield  {author} {\bibinfo {author} {\bibfnamefont {H.~M.}\ \bibnamefont
  {Vasconcelos}}, \bibinfo {author} {\bibfnamefont {L.}~\bibnamefont {Sanz}}, \
  and\ \bibinfo {author} {\bibfnamefont {S.}~\bibnamefont {Glancy}},\ }\href
  {\doibase 10.1364/OL.35.003261} {\enquote {\bibinfo {title} {All-optical
  generation of states for "encoding a qubit in an oscillator},}\ } (\bibinfo
  {year} {2010}),\ \bibinfo {note} {opt. Lett. 35, \textbf{3261}}\BibitemShut
  {NoStop}%
\bibitem [{\citenamefont {Bourassa}\ \emph {et~al.}(2021)\citenamefont
  {Bourassa}, \citenamefont {Alexander}, \citenamefont {Vasmer}, \citenamefont
  {Patil}, \citenamefont {Tzitrin}, \citenamefont {Matsuura}, \citenamefont
  {Su}, \citenamefont {Baragiola}, \citenamefont {Guha}, \citenamefont
  {Dauphinais}, \citenamefont {Sabapathy}, \citenamefont {Menicucci},\ and\
  \citenamefont {Dhand}}]{blueprint}%
  \BibitemOpen
  \bibfield  {author} {\bibinfo {author} {\bibfnamefont {J.~Eli}\ \bibnamefont
  {Bourassa}}, \bibinfo {author} {\bibfnamefont {R.~N.}\ \bibnamefont
  {Alexander}}, \bibinfo {author} {\bibfnamefont {M.}~\bibnamefont {Vasmer}},
  \bibinfo {author} {\bibfnamefont {A.}~\bibnamefont {Patil}}, \bibinfo
  {author} {\bibfnamefont {I.}~\bibnamefont {Tzitrin}}, \bibinfo {author}
  {\bibfnamefont {T.}~\bibnamefont {Matsuura}}, \bibinfo {author}
  {\bibfnamefont {D.}~\bibnamefont {Su}}, \bibinfo {author} {\bibfnamefont
  {B.~Q.}\ \bibnamefont {Baragiola}}, \bibinfo {author} {\bibfnamefont
  {S.}~\bibnamefont {Guha}}, \bibinfo {author} {\bibfnamefont {G.}~\bibnamefont
  {Dauphinais}}, \bibinfo {author} {\bibfnamefont {K.~K.}\ \bibnamefont
  {Sabapathy}}, \bibinfo {author} {\bibfnamefont {N.~C.}\ \bibnamefont
  {Menicucci}}, \ and\ \bibinfo {author} {\bibfnamefont {I.}~\bibnamefont
  {Dhand}},\ }\bibfield  {title} {\enquote {\bibinfo {title} {Blueprint for a
  scalable photonic fault-tolerant quantum computer},}\ }\href {\doibase
  10.22331/q-2021-02-04-392} {\bibfield  {journal} {\bibinfo  {journal}
  {Quantum}\ }\textbf {\bibinfo {volume} {5}},\ \bibinfo {pages} {392}
  (\bibinfo {year} {2021})}\BibitemShut {NoStop}%
\bibitem [{\citenamefont {Eaton}\ \emph {et~al.}(2019)\citenamefont {Eaton},
  \citenamefont {Rajveer},\ and\ \citenamefont {Olivier}}]{catalysis}%
  \BibitemOpen
  \bibfield  {author} {\bibinfo {author} {\bibfnamefont {M.}~\bibnamefont
  {Eaton}}, \bibinfo {author} {\bibfnamefont {N.}~\bibnamefont {Rajveer}}, \
  and\ \bibinfo {author} {\bibfnamefont {P.}~\bibnamefont {Olivier}},\
  }\bibfield  {title} {\enquote {\bibinfo {title} {Non-gaussian and
  gottesman--kitaev--preskill state preparation by photon catalysis},}\ }\href
  {\doibase 10.1088/1367-2630/ab5330} {\bibfield  {journal} {\bibinfo
  {journal} {New J. Phys.}\ }\textbf {\bibinfo {volume} {21}},\ \bibinfo
  {pages} {113034} (\bibinfo {year} {2019})}\BibitemShut {NoStop}%
\bibitem [{\citenamefont {Tzitrin}\ \emph {et~al.}(2020)\citenamefont
  {Tzitrin}, \citenamefont {Bourassa}, \citenamefont {Menicucci},\ and\
  \citenamefont {Sabapathy}}]{approx}%
  \BibitemOpen
  \bibfield  {author} {\bibinfo {author} {\bibfnamefont {I.}~\bibnamefont
  {Tzitrin}}, \bibinfo {author} {\bibfnamefont {J.~Eli}\ \bibnamefont
  {Bourassa}}, \bibinfo {author} {\bibfnamefont {N.~C.}\ \bibnamefont
  {Menicucci}}, \ and\ \bibinfo {author} {\bibfnamefont {K.~K.}\ \bibnamefont
  {Sabapathy}},\ }\bibfield  {title} {\enquote {\bibinfo {title} {Progress
  towards practical qubit computation using approximate
  gottesman-kitaev-preskill codes},}\ }\href {\doibase
  10.1103/PhysRevA.101.032315} {\bibfield  {journal} {\bibinfo  {journal}
  {Phys. Rev. A}\ }\textbf {\bibinfo {volume} {101}},\ \bibinfo {pages}
  {032315} (\bibinfo {year} {2020})}\BibitemShut {NoStop}%
\end{thebibliography}%

\end{document}